\documentclass[aps,onecolumn,pre,showpacs,balancelastpage,amssymb,groupedaddress]{revtex4}
\usepackage{graphicx}
\usepackage{amsmath}

\newcommand{\tr}{\operatorname{tr}}

\newcommand{\ei}{\mathrm{e}}
\newcommand{\bc}{\begin{center}}
\newcommand{\ec}{\end{center}}
\newcommand{\be}{\begin{equation}}
\newcommand{\ee}{\end{equation}}
\newcommand{\bq}{\begin{eqnarray}}
\newcommand{\eq}{\end{eqnarray}}

\begin{document}
\title{Characteristic Time and Maximum Mixedness: Single Mode Gaussian States in Dissipative Channels}
\author{Leonardo A. M. Souza\footnote{Corresponding author.} and M. C. Nemes}
\affiliation{Departamento de F\'{\i}sica, Instituto de
Ci\^{e}ncias Exatas, Universidade Federal de Minas Gerais, CP 702,
CEP 30161-970, Belo Horizonte, Minas Gerais, Brasil, e-mail: lamsouza@fisica.ufmg.br, Phone: +55-31-3499-5605, Fax: +55-31-3499-5688.}

\begin{abstract}

We derive an upper limit for the mixedness of single bosonic mode
gaussian states propagating in dissipative channels. It is a
function of the initial squeezing and temperature of the channel
only. Moreover the time at which von Neumann's entropy reaches its
maximum value coincides with that of complete loss of coherence,
thus defining a quantum-classical transition.



\end{abstract}

\maketitle

\section{Introduction} The search for quantum squeezed states of light began in the
1950's with the pioneering works of Senitzky, Plebanski and others
\cite{senitzky,husimi,plebanski,epstein}. The recent and rapid
development of quantum information has greatly stimulated research
on nonclassical states of light. They play an important role in
quantum information processing with continuous variables
\cite{braustein1}, in which information is encoded in two
conjugate quadratures of an optical field mode. Moreover, quantum
mechanical squeezing of optical fields represents a means to
improve the precision of measurements below the standard quantum
noise detection. Also, squeezed states may be used as an
entanglement source: for example, combining two squeezed states at
a beam splitter creates an entangled two-mode squeezed state, as
required for quantum teleportation \cite{bennett} or dense coding
schemes \cite{braustein2}. On the experimental side, a proposed
scheme \cite{jaromir1} for measuring squeezing, purity and
entanglement of Gaussian squeezed states (GSS) of light was
realized recently \cite{jaromir2}.

As is well known, dissipative environments tend to render quantum
features unobservable. Dissipative dynamics of quantum systems
have been investigated, both theoretically
\cite{caldeira1,caldeira,zurek,cavalcanti,lee,han,subpoissonian,manko,lages,rossi,saraceno,karen,knight1,oliveira,knight3,knight2,dodonov,ozorio,italianos1,italianos,marian2}
and experimentally
\cite{paris,paris1,walborn,experimental,experimental1}.
Understanding the factors which limit the visibility and also the
time scales of quantum properties (QP) is of interest if they are
to be used in the context of quantum computation or foundations of
quantum theory, e.g. the quantum - classical border.


In the present contribution we analytically derive two essential
ingredients for the visibility of QP (in the context of to a
Caldeira-Legget type environment): (i) an upper limit for the
mixedness of the initial single mode bosonic GSS and (ii) a time
scale for (the observables) squeezing and oscillations in photon
number distributions. These two conditions, considered together,
convey in simple terms the relevant control parameters in an
experiment involving such states.



\section{Single mode bosonic Gaussian States and their dissipative dynamics}

In this work we deal with single mode displaced, squeezed, mixed
Gaussian States. This class of states can always be put in the
form  \be \label{gaussini} \hat{\rho}_G = \mathcal{D}(\alpha_0)
\mathcal{S}(r_0,\phi_0) \hat{\rho}_{\nu_0}
\mathcal{S}^\dagger(r_0, \phi_0) \mathcal{D}^\dagger (\alpha_0)
\ee where $\mathcal{D}(\alpha)$ is the displacement operator,
$\mathcal{S}(r,\phi)$ is the squeezing operator and
$\hat{\rho}_\nu$ is the thermal density operator whose degree of
mixedness is given by  $\tr [a^\dagger a \rho_\nu] = \nu$. The
subscripts {0} denote initial values. More explicitly we have \bq
\mathcal{D}(\alpha) &=& \exp(\alpha a^\dagger - \alpha^* a) \\
\mathcal{S}(r,\phi) &=& \exp \Big{(} \frac{1}{2} r \ei^{i \phi}
a^{\dagger 2} - \frac{1}{2} r \ei^{-i \phi} a^2 \Big{)}
\\ \hat{\rho}_\nu &=& \frac{1}{1+\nu} \exp \Bigg{(} \ln \Big{(}
\frac{\nu}{\nu+1} \Big{)} a^\dagger a \Bigg{)}. \eq

The dissipative dynamics is taken to be the following master
equation (we will work in units such that $\hbar = 1$) \be
\label{master} \dot{\rho} = \mathcal{L} \rho \ee where
$\mathcal{L}$ is the Liouvillian super-operator : \bq \mathcal{L}
&=& - i \omega [a^\dagger a , \bullet] + k (\bar{n}_B + 1) (2 a
\bullet a^\dagger - a^\dagger a \bullet - \bullet a^\dagger a) +
\nonumber
\\ && + k ~ \bar{n}_B (2 a^\dagger \bullet a - a a^\dagger \bullet -
\bullet a a^\dagger), \eq known to yield good quantitative
agreement with  data from quantum optics (the $\bullet$ indicates
where the super-operator acts). In equation \eqref{master}
$\omega$ stands for the field frequency, $\bar{n}_B$ is the
average number of thermal photons and $k$ the dissipation
constant.

The analytical solution of \eqref{master} for states of the form
\eqref{gaussini} are given by \cite{serafini} \bq
\label{alpha1} \alpha (t) &=& \alpha_0 \ei^{-(i \omega + k) t} \\
\phi(t) &=& \phi_0 - 2 \omega t \\ \nu(t) &=& \sqrt{x^2(t) -
\Big{(}(\nu_0+\frac{1}{2})
\sinh (2 r_0) \ei^{-2 k t} \Big{)}^2}-\frac{1}{2} \nonumber \\
\\x(t) &=& \Big{(} \nu_0 + \frac{1}{2}\Big{)} \cosh(2 r_0) \ei^{-2 k
t} + \Big{(} \overline{n}_B + \frac{1}{2} \Big{)} (1-\ei^{-2 k t})
\nonumber \\ \\ \label{rt} r(t)&=& \frac{1}{4} \ln \Bigg{(}
\frac{(\nu_0 + \frac{1}{2}) \ei^{2 r_0}+ (\overline{n}_B +
\frac{1}{2}) (\ei^{2 k t} -1) }{(\nu_0 + \frac{1}{2}) \ei^{-2
r_0}+ (\overline{n}_B + \frac{1}{2}) (\ei^{2 k t} -1) } \Bigg{)}
\nonumber.
\\ \eq

\begin{figure} [!h]
\begin{center}
\caption{von Neumann entropy for a GSS with $r_0 = 1$, $k = 0.1$
and $\bar{n}_B=0$. For the initial mixedness of the quantum state
we have: $\nu_0 = 3$ the dashed curve (left scale); $\nu_0 = 0$
the solid curve (right scale).} \label{figure1}
\includegraphics [width = 7cm,height=7cm,angle=0] {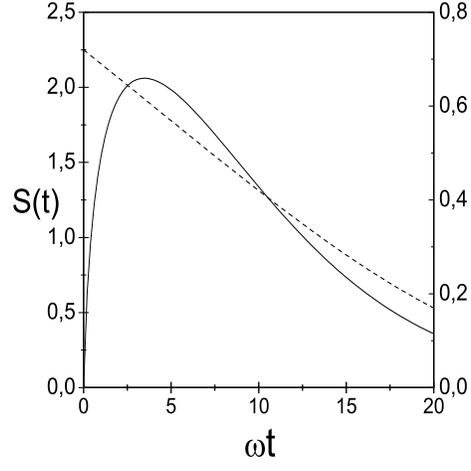}
\end{center}
\end{figure}

Now we use a well known property of Gaussian states that allows us
to obtain an analytical expression for von Neumann's entropy
\cite{agarwal}: any Gaussian state is completely determined by the
mean values of position and momentum operators as well as their
quadratures. The property we have in mind is the determinant of
the covariance matrix
\begin{align} \label{det}
  D(t) &= \textrm{Det}
\begin{pmatrix}
    \sigma_{pp} & \sigma_{qp}\\
    \sigma_{qp} & \sigma_{qq}
  \end{pmatrix}
\end{align} where $\sigma_{xy}= \frac{1}{2} \tr [{\hat
\rho}\left\{\hat{x},\hat{y}\right\}] -\tr[\hat{x}\hat
\rho]\tr[\hat{y}\hat \rho]$. It is easy to show that, for a single
mode GSS: \be \label{det} D(t) = \Big{(} \nu(t) + \frac{1}{2}
\Big{)}^2 \ee and \be \label{entropy} S[\rho_\nu(t)] = (\nu(t)+1)
\ln \big{(} \nu(t)+1 \big{)} - \nu(t) \ln \big{(} \nu(t) \big{)}
\ee where $S[\rho_\nu(t)] = - \tr [\rho_\nu \ln \rho_\nu]$ is the
von Neumann entropy. Note that the coherence content of the time
evolved GSS is \emph{independent} of the displacement $\alpha(t)$
and the squeezing phase $\phi(t)$. The QP contained in a GSS are
its squeezing and oscillations in  photon number distribution.
During the time interval when such properties are visible one
finds an increase in entropy which afterwards decays to the purely
Gaussian vacuum state (for zero temperature). In figure
\eqref{figure1} we illustrate how the initial mixedness of the GSS
influences the visibility of quantum coherence. It shows the
entropy for a pure ($\nu_0 = 0$) and a mixed Gaussian state
($\nu_0 = 3$). In the present case the time scale for both
squeezing and oscillations in photon number distributions
coincides with the time at which the von Neumann entropy attains
its maximum value.

%

A physical understanding of the dynamical process is best afforded
by the Wigner function (WF). For a GSS we get \bq W(x,p) &=&
\sum_l \frac{1}{\pi} \frac{\nu^l}{(\nu+1)^{l+1}} (-|F_3|)^l \cdot
L_l \Big{[} 2 \Big{(} \frac{(x-x_0)^2}{F_4^2} + \frac{(F_4
F_5)^2}{4} \Big{)}\Big{]} \cdot \nonumber \\ && \cdot
\frac{F_4}{|F_1|} \ei^{-\frac{(x- x_0)^2}{F_4^2}} \ei^{-\frac{(F_4
F_5)^2}{4}}, \eq where $L_l(x)$ is the Laguerre function of $l$
order (and argument $x$) and we define \cite{nieto} \bq F_1 &=&
\cosh r + \ei^{i \phi} \sinh r \nonumber
\\ F_2 &=& \frac{1- i \sin \phi \sinh r (\cosh r + \ei^{i \phi}
\sinh r)}{(\cosh r + \cos \phi \sinh r)(\cosh r + \ei^{i \phi}
\sinh r)} \nonumber \\ F_3 &=& \frac{\cosh r + \ei^{-i \phi} \sin
\phi \sinh r}{\cosh r + \ei^{i \phi} \sin \phi \sinh r} \nonumber \\
F_4 &=&\sqrt{\cosh^2 r + \sinh^2 r + 2 \cos \phi \cosh r \sinh r}
\nonumber\\ F_5 &=& 2 (p+p_0)- i (x-x_0)(F_{2}^{*}-F_2).  \eq 
The Wigner function above can also be written  in a closed
Gaussian form (since the dynamics do not change its Gaussian
character): \bq W(x,p) &=& \frac{1}{\pi (\nu + \frac{1}{2})} \exp
\Big{\{} - \frac{\cosh(2 r)}{2 \nu +1} \left[ (1-\tanh(2 r) \cos
\phi) x^2 + (1+\tanh(2 r) \cos \phi) p^2 \right] + \nonumber \\ &&
+ \frac{\sin \phi \sinh(2 r)}{\nu + \frac{1}{2}} x p \Big{\}}. \eq
At time zero the WF shows apparent squeezing in $x$, afterwards
the WF rotates in the $x-p$ plane and tends to become a pure
Gaussian state. Its maximum becomes lower, the width in the
initially squeezed quadrature increases and decreases in the
other. When these squeezing effects disappear the entropy attains
its maximum value, and behaves very similarly to the case where
$\nu_0 = 3$ (keeping the other parameters constant). In this case,
there is an initial squeezing also, and the same dynamics takes
place, however the observation time for these properties {\it is
practically} zero.



\begin{figure} [!h]
\caption{Photon number distribution for a GSS with $r_0 = 1$, $k =
0.1$, $\nu_0=0$ and $\bar{n}_B=0$.} \label{figure3}
\begin{center} \vspace{0cm} 
\includegraphics [scale=0.45] {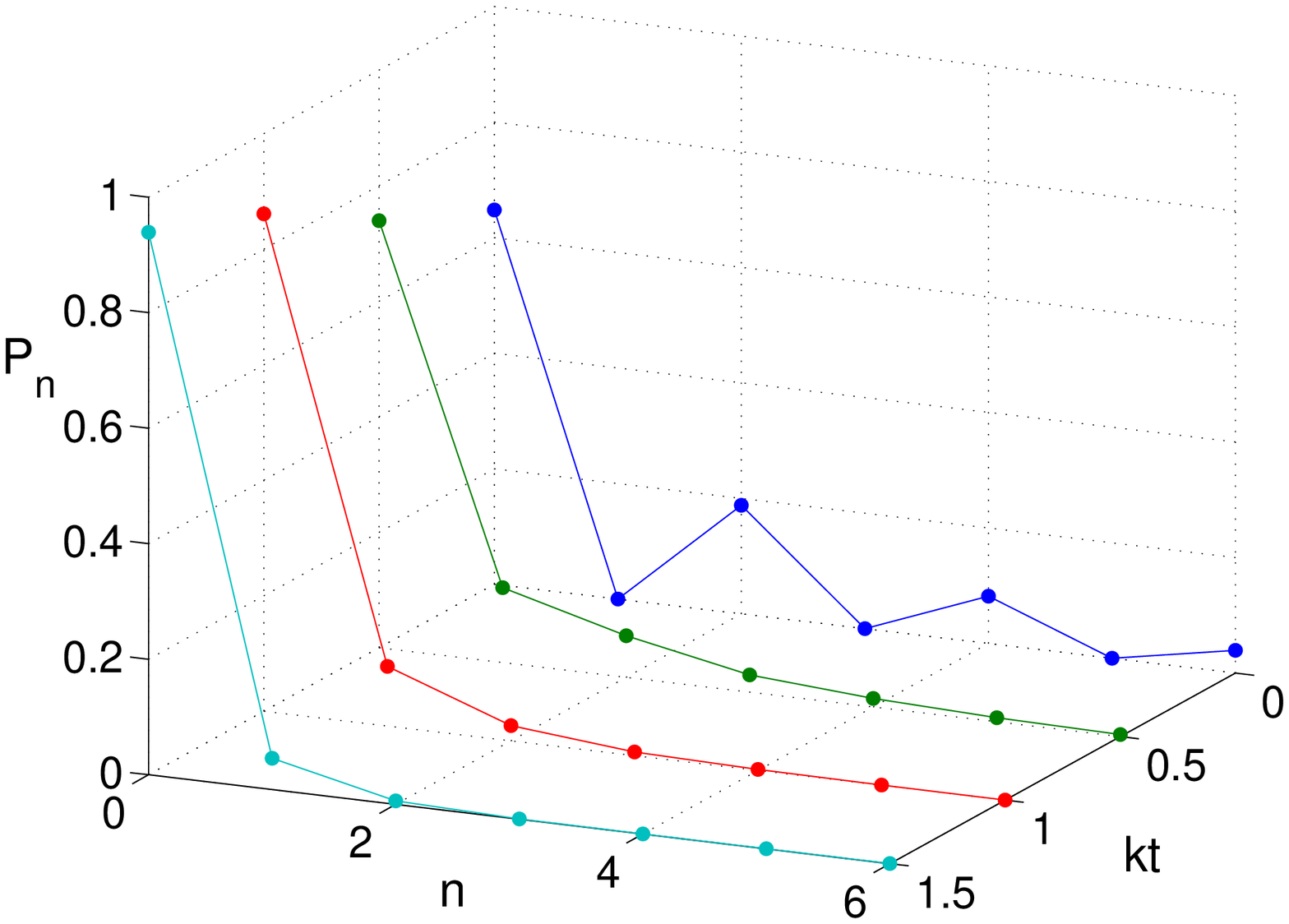}
\end{center} 
\end{figure}

\begin{figure} [!h]
\caption{Photon number distribution for a GSS with $r_0 = 1$, $k =
0.1$, $\nu_0=3$ and $\bar{n}_B=0$.} \label{figure4}
\begin{center} \vspace{0cm} 
\includegraphics [scale=0.45] {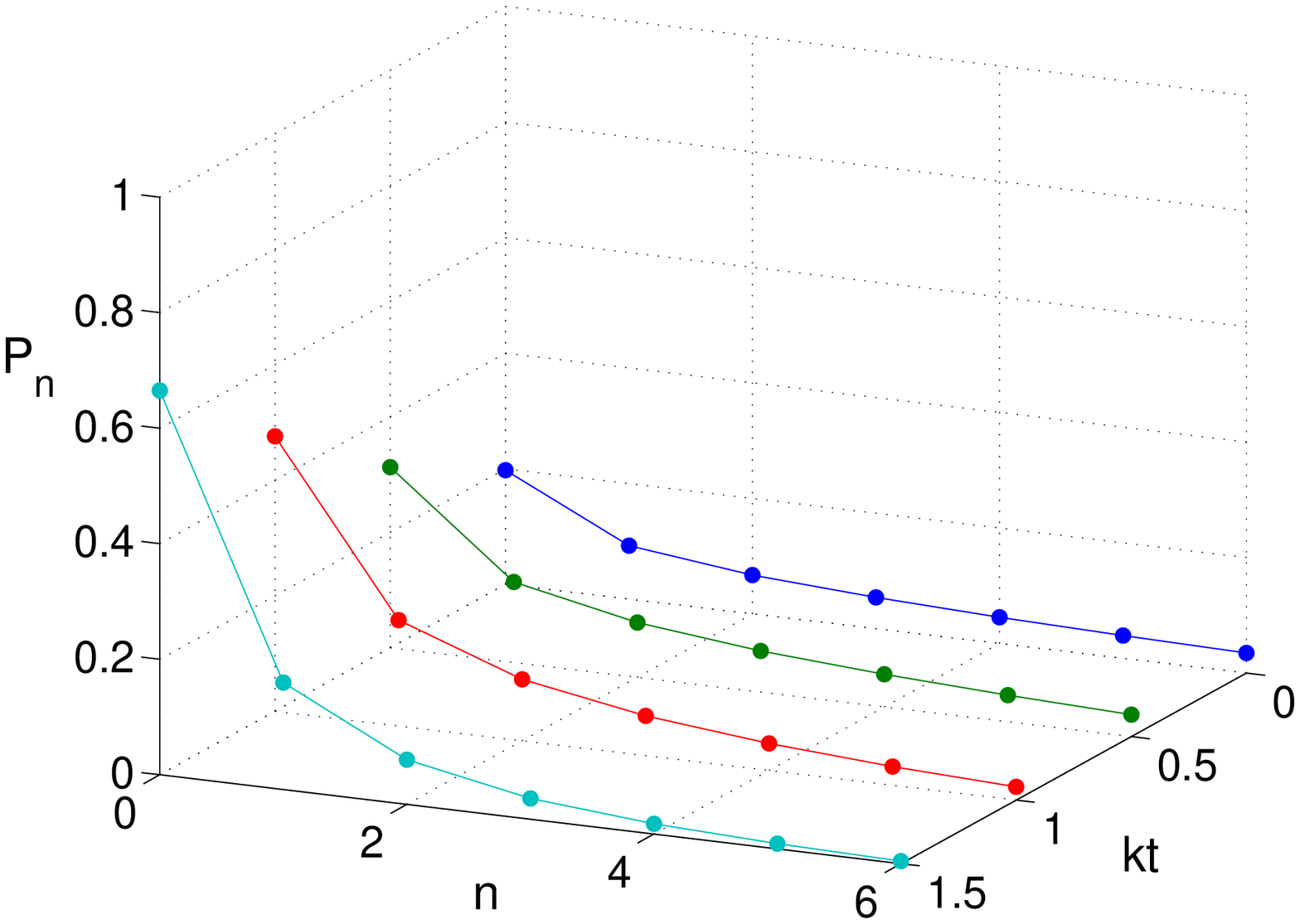}
\end{center}
\end{figure}

Next we study the photon number distribution which is
given by \cite{marian,marian1} \bq \rho_{n n} &=& P_n = \pi Q(0) (-1)^n
2^{-2 n} (\tilde{A}+|\tilde{B}|)^n \nonumber \\ && \times
\sum_{k=0}^{n} \frac{1}{k! (n-k)!}\Big{[} \frac{\tilde{A}-
|\tilde{B}|}{\tilde{A}+|\tilde{B}|} \Big{]}^k \nonumber
\\ && \times H_{2 k} \Bigg{[} i \frac{\Im (\tilde{C} \ei^{-i
\frac{\phi}{2}})}{\sqrt{\tilde{A}-|\tilde{B}|}} \Bigg{]} \nonumber \\
&& \times H_{2 n - 2 k} \Bigg{[} i \frac{\Re (\tilde{C} \ei^{-i
\frac{\phi}{2}})}{\sqrt{\tilde{A}+|\tilde{B}|}} \Bigg{]} \eq where
$H_j$ is the j-order Hermite polynomial and \bq \pi Q(0) &=&
[(1+A)^2-|B|^2]^{1/2} \nonumber \\ && \cdot \exp \Bigg{\{} -
\frac{(1+A) |C|^2 + \frac{1}{2} [B(C^*)^2+B^* C^2]}{(1+A)^2-|B|^2}
\Bigg{\}}.\eq  and \bq A &=& \nu + (2 \nu +1) \sinh^2 r
\\ B &=& -(2 \nu +1) \ei^{i \phi} \sinh r \cosh r \\ C &=& \alpha
\\ \tilde{A} &=& \frac{\nu (\nu + 1)}{\nu^2 + (\nu +
\frac{1}{2})[1+\cosh (2 r)]}
\\ \tilde{B} &=& - \frac{\ei^{i \phi} (\nu \frac{1}{2}) \sinh (2
r)}{\nu^2 + (\nu + \frac{1}{2})[1+\cosh (2 r)]} \\ \tilde{C} &=&
\frac{C [\frac{1}{2} + (\nu + 1/2) \cosh (2 r)]- C^* \ei^{i \phi}
(\nu + \frac{1}{2}) \sinh (2 r)}{\nu^2 + (\nu + \frac{1}{2})[1+
\cosh (2 r)]}.\eq Note that $\nu$, $\phi$, $\alpha$ and $r$ are
\emph{time dependent}. Figures \eqref{figure3} and \eqref{figure4}
show the result corresponding to the time evolution for the
initial conditions of figure \eqref{figure1} and for $\nu_0=0$ and
$\nu_0 = 3$ respectively. One can see that for suitable initial
conditions oscillations in photon number are visible for a given
amount of time. As the degree of mixedness is increased (figure
\eqref{figure4}), leaving the other parameters unchanged, only
thermal-like (classical) distributions are seen.



In the case of 1-mode GSS,  there is only one time scale in the
problem, that given by the time at which the entropy attains its
maximum value, i. e., the decoherence time, given by (this result
was also found in ref. \cite{marian1}, for the linear entropy) \bq
\label{tcgauss} t_{c} &=& (2 k)^{-1} \Big{\{} \ln 2 -
\ln \Big{[} \frac{2 \bar{n}_B +1}{d}   \\
&\times&(2 \nu_0 \cosh (2 r_0) + \cosh (2 r_0) -2 \bar{n}_B -1)
\Big{]} \Big{\}}\nonumber\eq where \bq d &=&
2\cosh\left(2r_0\right) \left[\bar{n}_B\left(\nu_0+1\right) +
\nu_0\left(\bar{n}_B + 1\right)+\frac{1}{2}\right] \nonumber \\
&&-2\left(\bar{n}_B+\frac{1}{2}\right)^2-2\left(\nu_0+\frac{1}{2}\right)^2.
\nonumber \eq We obtain this characteristic time by studying the
time evolution of $D(t)$ (or the entropy). Explicitly $t_c$ is the
time when $D(t)$ reaches its maximum value, i.e., we define $t_c$
as the positive solution for $t$ of: \be \frac{\partial}{\partial
t} D(t) = 0. \ee We remark that, for consistency, when $t_c < 0$
we assume the characteristic time to be zero, i.e., the system
simply tends to equilibrium with the environment, without any
``quantum dynamics''.

After discussing those QP of GSS we show the main result of this
contribution: quantum properties will still be visible in
initially Gaussian states provided \be \label{inequal} \nu_0 <
\cosh(2 r_0)\Big{(}\bar{n}_B+\frac{1}{2}\Big{)}-\frac{1}{2}.\ee
This relation was obtained from analysis of equation \eqref{det},
where we studied when and in what conditions the function reaches
its maximum. As a matter of fact, there must be a compromise
between $\nu_0$ and the temperature for QP to be visible, i.e.,
{\it given an initial condition $\nu_0$}, the temperature must
satisfy \be \bar{n}_B < \cosh(2
r_0)\Big{(}\nu_0+\frac{1}{2}\Big{)}-\frac{1}{2},\ee such that an
increase in $D(t)$ (or in the entropy) is possible. Of course,
aside from obeying the above inequalities the time scales must be
experimentally accessible.

Let us analyze the case with $\bar{n}_B = 0$ for simplicity
without loss of generality. There is a clear competition between
squeezing and the degree of purity of the initial state: if the
squeezing parameter is large enough quantum properties may still
be visible for nonpure states with a given degree of mixedness. We
have thus show that a growth in entropy signal the presence of
squeezing and/or oscillations in the photon number distribution.

\section{Conclusions.} In summary, we reviewed some quantum characteristics
of single mode mixed Gaussian states such as oscillations in
photon number distribution and squeezing undergoing a dissipative
dynamics and obtained an upper limit for their initial degree of
mixedness such that quantum properties may still be visible,
provided the time scale for the process in question is realistic
from the experimental view point.

~
~

{\bf Acknowledgments} L. A. M. Souza thanks CNPq-Brasil for
financial support. M. C. Nemes was partially supported by
CNPq-Brasil. The authors would like to thank Professor R. Dickman
for the suggestions about the paper.

\newpage

\begin{center} Figure captions \end{center}

\center{FIG. 1: von Neumann entropy for a GSS with $r_0 = 1$, $k =
0.1$ and $\bar{n}_B=0$. For the initial mixedness of the quantum
state we have: $\nu_0 = 3$ the dashed curve (left scale); $\nu_0 =
0$ the solid curve (right scale).}

\center{FIG. 2: Photon number distribution for a GSS with $r_0 =
1$, $k = 0.1$, $\nu_0=0$ and $\bar{n}_B=0$.}

\center{FIG. 3: Photon number distribution for a GSS with $r_0 =
1$, $k = 0.1$, $\nu_0=3$ and $\bar{n}_B=0$.}

\newpage

\newpage

\newpage

\begin{thebibliography}{99}

\bibitem{senitzky}

I. R. Senitzky, Phys. Rev. \textbf{95}, 1115 (1954).

\bibitem{plebanski}

J. Plebanski, Phys. Rev. \textbf{101}, 1825 (1956).

\bibitem{husimi}

K. Husimi, Prog. Theor. Phys. \textbf{9}, 381 (1953).

\bibitem{epstein}

S. T. Epstein, Am. J. Phys. \textbf{27}, 291 (1959).

\bibitem{braustein1}

S. L. Braunstein and P. van Loock, Rev. Mod. Phys. \textbf{77},
513 (2005).

\bibitem{bennett}

C. H. Bennett et al, Phys. Rev. Lett. \textbf{70}, 1895 (1993).

\bibitem{braustein2}

S. L. Braunstein and H. J. Kimble, Phys. Rev. A \textbf{61}, 042302
(2000).

\bibitem{jaromir1}

J. Fiur\'{a}\^{s}ek and N. J. Cerf, Phys. Rev. Lett. \textbf{93},
063601 (2004).

\bibitem{jaromir2}

J. Wenger et al, Phys. Rev. A \textbf{70}, 053812 (2004).

\bibitem{caldeira1}

A. O. Caldeira and A. J. Leggett, Phys. Rev. Lett. \textbf{46},
211 (1981).

\bibitem{caldeira}

A. O. Caldeira and A. J. Leggett, Ann. Phys. (N.Y.) \textbf{149},
374 (1983).

\bibitem{zurek}

R. Blume-Kohout and W. H. Zurek, Phys. Rev. A \textbf{68} 032104
(2003).

\bibitem{cavalcanti}

E. G. Cavalcanti and M. D. Reid, Phys. Rev. Lett. \textbf{97},
170405 (2006).

\bibitem{lee}

J. W. Lee and D. L. Shepelyansky, Phys. Rev. E \textbf{71}, 056202
(2005).

\bibitem{han}

H. Han et al, J. Phys. B: At. Mol. Opt. Phys. \textbf{40}, S209
(2007).

\bibitem{subpoissonian}

G. Y. Kryuchkyan and S. B. Manvelyan, Phys. Rev. A \textbf{68},
013823 (2003).

\bibitem{manko}

V. V. Dodonov, O. V. Man'ko and V. I. Man'ko, Phys. Rev. A
\textbf{49}, 2993 (1994).

\bibitem{lages}

J. Lages et al, Phys. Rev. E \textbf{72}, 026225 (2005).

\bibitem{rossi}

R. Rossi Jr., A. R. B. Magalh\~{a}es and M. C. Nemes, Phys. Lett.
A \textbf{356}, 277 (2006).

\bibitem{saraceno}

P. Bianucci, J. P. Paz and M. Saraceno, Phys. Rev E \textbf{65},
046226 (2002).

\bibitem{karen}

K. M. F. Romero and M. C. Nemes, Physica A \textbf{325}, 333
(2003).

\bibitem{knight1}

K. W\'{o}dkiewicz et al, Phys. Rev A \textbf{35}, 2567 (1987).

\bibitem{oliveira}

M. S. Kim, F. A. M. de Oliveira and P. L. Knight, Phys. Rev. A
\textbf{40}, 2494 (1989); F. A. M. de Oliveira et al, Phys. Rev. A
\textbf{41}, 2645 (1990).

\bibitem{knight3}

V. Bu\v{z}ek, A. Vidiella-Barranco and P. L. Knight, Phys. Rev. A
\textbf{45}, 6570 (1992).

\bibitem{knight2}

L. Gilles and P. L. Knight, Phys. Rev. A \textbf{48}, 1582 (1993).

\bibitem{dodonov}

A. S. M. de Castro and V. V. Dodonov, Phys. Rev. A \textbf{73},
065801-1 (2006).

\bibitem{ozorio}

A. M. O. de Almeida, J. Phys. A.: Math. Gen. \textbf{36}, 67
(2003).

\bibitem{italianos1}

Matteo G. A. Paris et al, Phys. Rev. A \textbf{68}, 012314 (2003).

\bibitem{italianos}

Serafini A., Paris M. G. A., Illuminati F. and De Siena
 S., J. Opt. B: Quantum Semiclass. Opt. \textbf{7}, R19 (2005).

\bibitem{marian2}

P. Marian and T. A. Marian, J. Phys. A: Math. Gen. \textbf{33}, 3595 (2000).

\bibitem{paris}

M. Brune et al, Phys. Rev. A \textbf{45}, 5193 (1992).

\bibitem{paris1}

M. Brune et al, Phys. Rev. Lett. \textbf{65}, 976 (1990).

\bibitem{walborn}

S. P. Walborn et al, Nature \textbf{440}, 1022 (London, 2006).

\bibitem{experimental}

P. Sonnentag and F. Hasselbach, Phys. Rev. Lett. \textbf{98},
200402 (2007).

\bibitem{experimental1}

W. Wang, L. B. Fu, and X. X. Yi, Phys. Rev. A \textbf{75}, 045601
(2007).

\bibitem{serafini}

A. Serafini et al, Phys. Rev. A \textbf{69}, 023318 (2004).

\bibitem{agarwal}

G. S. Agarwal, Phys. Rev. A \textbf{3}, 828 (1971); for a review
about Robertson-Schr\"{o}dinger uncertainty principle see
arXiv:quant-ph/9903100v2.

\bibitem{nieto}

M. M. Nieto, Phys. Lett. A \textbf{229}, 135 (1997).

\bibitem{marian}

P. Marian and T. A. Marian, Phys. Rev. A \textbf{47}, 4474 (1993).

\bibitem{marian1}

P. Marian and T. A. Marian, Phys. Rev. A \textbf{47}, 4487 (1993).


\end{thebibliography}
\end{document}